\documentclass[twocolumn,showpacs,amsmath,amssymb,prd,nofootinbib]{revtex4}
\usepackage{epsfig}
\def\sss{\scriptscriptstyle}
\def\^#1{^{\sss #1}}
\def\_#1{_{\sss #1}}
\def\beq{\begin{equation}}
\def\eeqno#1{\label{#1}\end{equation}}

\def\kms{{\rm km~s^{-1}}}

\def\kpc{{\rm Kpc}}

\def\msun{M\_{\odot}}
\def\az{a\_{0}}

\def\l0{\ell\_{0}}

\def\l{\lambda}

\def\m{\mu}

\def\a{\alpha}
\def\b{\beta}

\def\gN{g\_N}

\begin{document}
\title{The $\Lambda$CDM simulations of Keller and Wadsley  do not account for the MOND mass-discrepancy-acceleration relation}
\author{Mordehai Milgrom}
\affiliation{Department of Particle Physics and Astrophysics, Weizmann Institute\\and\\Institut d'Astrophysique de Paris}

\begin{abstract}
Keller and Wadsley (2016) have smugly suggested, recently, that the end of MOND may be in view. This is based on their claim that their highly-restricted sample of $\Lambda$CDM-simulated galaxies are ``consistent'' with the observed MOND mass-discrepancy-acceleration relation (MDAR); in particular, with its recent update by McGaugh et al. (2016), based on the SPARC sample. From this they extrapolate to ``$\Lambda$CDM is fully consistent'' with the MDAR.
I explain why these simulated galaxies do not show that $\Lambda$CDM accounts for the MDAR.
a. Their sample of simulated galaxies contains only 18 high-mass galaxies, within a narrow range of one order of magnitude in baryonic mass, at the very high end of the observed, SPARC sample, which spans 4.5 orders of magnitude in mass. More importantly, the simulated sample has none of the low-mass, low-acceleration galaxies -- abundant in SPARC -- which encapsulate the crux and the nontrivial aspects of the predicted and observed MDAR. The low-acceleration part of the simulated MDAR is achieved, rather trivially, from the flattish-velocity-curve regions of the simulated high-mass galaxies.
b. Half of the simulated galaxies have ``wrong'' rotation curves that differ greatly from any observed ones. This, does not prevent these wrong galaxies from lying on the observed MDAR (for trivial reasons, again). They, in fact, define the high-acceleration branch of the simulated MDAR.
c. To boot, even if $\Lambda$CDM were made ``consistent'' with the MDAR through the elaborate adjustments that the simulations allow, this would not obviate MOND, which predicts much more than the MDAR.

\end{abstract}
\maketitle

\section{\label{introduction} Introduction}
In a paper entitled `` La Fin du MOND? $\Lambda$CDM is Fully Consistent with SPARC'', Keller and Wadsley \cite{kw16} recently announced the possible end of MOND. This is because -- so they claim -- $\Lambda$CDM produces (simulated) galaxies that are ``consistent'' with the MOND mass-discrepancy-acceleration relation (MADR), recently updated and improved in Ref. \cite{mcgaugh16}.
\par
The MDAR emerges when one uses many observed rotation curves of disc galaxies, $V(r)$, to plot the measured accelerations at various radii, $g(r)=V^2(r)/r$, against the Newtonian, gravitational acceleration produced by baryons at that radius, $\gN(r)$. MOND has predicted \cite{milgrom83} that this should not produce a scatter plot with much scatter, as expected in the dark-matter paradigm, but that $g$ should be tightly correlated with $\gN$. Indeed, a large class of MOND theories predict that $g$ is a function of $\gN$ \cite{milgrom94}. This tight correlation is the MDAR. MOND also predicted that the MDAR has two branches, connected smoothly, the transition occurring around $\gN=\az$, where $\az$ is the MOND acceleration constant: For $\gN\gg\az$ MOND predicted $g\approx \gN$ -- the Newtonian, or high-acceleration branch -- while for $\gN\ll\az$ MOND predicted $g\approx (\az\gN)^{1/2}$ -- the deep-MOND, or low-acceleration branch.
\par
These predictions have been amply vindicated by many studies in the past (see references listed in Ref. \cite{milgrom16a}), culminating in the recent analysis of Ref. \cite{mcgaugh16}, using their SPARC sample of disc galaxies.

\par
Reference \cite{kw16} shows that a highly-restricted sample of their own $\Lambda$CDM-simulated galaxies lie on the observed MDAR at redshift $z=0$ (not at higher $z$). They then jump to the conclusion  that $\Lambda$CDM is ``fully consistent'' with the MDAR. And so, by further unwarranted extrapolation, they seem to imply that $\Lambda$CDM is consistent with all the observed galaxy properties predicted by MOND, and that this may mean that there is no need for MOND.
\par
It is, however, nothing of the sort: The analysis of Ref. \cite{kw16} does not demonstrate that $\Lambda$CDM
accounts for the observed MDAR. And, even if simulated galaxies fall on the MDAR it does not mean that they are `correct' galaxies (many of them are not). And, even if the MDAR is consistent with $\Lambda$CDM, this does not obviate MOND.
\par
It also befits us to remember that while MOND predicted the MDAR over thirty years ago, $\Lambda$CDM cannot predict such relations, which result from complicated baryonic processes.
\par
In Sec. \ref{mond}, I explain why there is much more to MOND than the MDAR. In Secs. \ref{analysis} and \ref{wrong}, I discuss the limitation of the analysis of Ref. \cite{kw16}, and explain why it does not show that $\Lambda$CDM is ``fully consistent'' with the MDAR. I make some further comments in Sec. \ref{comments}.
\section{\label{mond}MOND is more than the MDAR}
I will start with the last point.
MDAR, which pertains to rotation curves of disc galaxies, is only part of the successful predictions of MOND \cite{milgrom83} (see reviews in Refs. \cite{fm12,milgrom14c} for the various MOND predictions and how they fair). For example, there are the MOND predictions for dwarf spheroidal satellites (e.g., Refs. \cite{sad10,mm13a,mm13b}), for elliptical galaxies (e.g., Ref. \cite{milgrom12} and references therein), motions outside the planes of disc galaxies, as probed, e.g. by weak lensing by all galaxy types \cite{milgrom13}, etc. The discussion of Ref. \cite{kw16} concerns none of these.
\par
Even for rotation curves of disc galaxies, which underlie the MDAR, there are the very quintessential MOND predictions, and their vindication, concerning the many so-called (very) low-surface-brightness, or low-acceleration disc galaxies.
This includes the mass-asymptotic-speed relation (underlying the observed baryonic-Tully-Fisher relation) as extended to the low-mass end.
\par
All these are wholly outside the treatment of Ref. \cite{kw16}, as their sample does not include any relevant, simulated galaxies (see Sec. \ref{analysis}).

\par
Also notable for this argument is the MOND prediction \cite{milgrom09b,milgrom16}, for disc galaxies, of a tight correlation between the central surface density of the baryonic disc, and the central, dynamical surface density, recently confirmed in Ref. \cite{lelli16}. Especially poignant is the predicted low-surface-density branch of this correlation, which is not even accessible by the sample of Ref. \cite{kw16}.

\par
Thus, Ref. \cite{kw16} does not, and cannot, show that any of the above MOND predictions are accounted for by $\Lambda$CDM simulations.

\section{\label{analysis}The Keller and Wadsley amalysis}
Coming now to the MDAR itself.
The MDAR plotted by Ref. \cite{mcgaugh16} is based on their SPARC sample of 153 observed galaxies with measured rotation curves and $3.6 \m$ photometry. Their sample (see their Fig. 1) spans a gamut of 4.5 orders of magnitude in baryonic mass $10^7\msun$ to $5\times 10^{11}\msun$. Perhaps more importantly, the sample span some 2.5 orders of magnitude in surface brightness, and includes many galaxies that are of very low acceleration (in the MOND sense) everywhere in their bulk.\footnote{high mass is correlated positively with high surface density.}
\par
In stark contrast with the SPARC sample, the simulated sample of Ref. \cite{kw16} involves only ``18 cosmological zoom-in simulations of $L*$ disc galaxies''. Their baryonic masses are all within the relatively narrow range of $\sim (0.17-2.7)\times 10^{11}\msun$, overlapping with just the highest decade out of the 4.5 decades in the SPARC sample.
\par
The properties of the 18 simulated galaxies -- their masses, rotation curves, etc. -- are described in Ref. \cite{kwc16}, by the authors of Ref. \cite{kw16}.
\par
None of the simulated galaxies correspond to observed low-accelerations (in the MOND sense), and they reach accelerations much lower than $\az$ only in their asymptotic, flattish parts. This is not a mere quantitative fault. It has very important qualitative implications in the context of MOND. As stated above, this means that the simulated sample cannot even begin to mimic some of the important MOND predictions, and, in particular, it is very important in the context of the MDAR:
\par
I first discuss the all-important, law-acceleration branch of the MDAR -- where the mass discrepancies are large and MOND enters in full strength. The high-acceleration branch will be discussed in Sec. \ref{wrong}.

The low-acceleration branch of the observed MDAR picks up two independent types of contributions (see, e.g., Refs. \cite{mcgaugh16,milgrom16a}): One, which might be considered the more trivial, comes from regions on the asymptotic, constant-rotational-speed regions of rotation curves. This is ``trivial'' in the sense that it adds nothing to what is  already contained in asymptotic flatness, and the mass-asymptotic-speed relation (MASSR), vindicated by a specific version of the ``baryonic Tully-Fisher relation'' (BTFR). This is because $MG\az=V^4$ implies $g\equiv V^2/R=(\az \gN)^{1/2}$, which is the MOND, low-acceleration branch.
\par
The highly non-trivial aspect of the MDAR is that the low-surface-density galaxies, where accelerations are low everywhere in the galaxy, fall on the same line as the asymptotic rotation curves of all galaxies, independent of the complex and varied mass distributions. The low-accelerations in these two regimes have nothing to do with each other, outside the framework of MOND.
\par
This non-trivial aspect of the low-acceleration MDAR is out of reach for the sample of Ref. \cite{kw16}.
Instead, looking at the simulated rotation curves in Fig. 4 of ref. \cite{kwc16}, one can gather that the contributions to this branch in their analysis come almost entirely from the asymptotic, flattish regions of their rotation curves. The fact that these fall on the MDAR can be traced to the fact that such flattish regions are there, and that the velocities there where probably designed to obey some approximately correct MASSR. For example, if we have $V(r)\approx V\_{flat}\propto M^{1/\a}$, with $\a$ for these high-mass galaxies around the observed value of $\a\approx 4$,  then we have in these regions $g=V^2/r\propto M^\b\gN^{1/2}$, with $\b=(4-\a)/2\a$. With $M$ varying only within one order of magnitude, and $\a\approx 4$, the mass dependence produces very little scatter around the $g\propto \gN^{1/2}$ branch.\footnote{A roughly correct normalization of the simulated MASSR would also give the correct normalization.}
\par
This, I believe, is the trivial reason for the simulated galaxies populating the low-acceleration branch of the observed MDAR.
\par
The fact that the simulated rotation curves are asymptotically flattish, and that they obey an approximately correct MASSR for this restricted set of high-mass galaxies might be considered significant. But, a. it is much less than what is encapsulated in the full low-acceleration branch of the MASSR, and, b. it could be a result of various adjustments in the simulations over the years, which tended to make them look, in some restricted regards, like observed galaxies.
In other words, if someone claimed that he managed to produce $\Lambda$CDM simulations of high-mass galaxies that satisfy a MASSR approximately consistent with observation, it would not be considered much of an achievement.

\section{\label{wrong}Simulated galaxies can be, and are, wrong, even if they lie on the MDAR }

Simulated galaxies can be wrong in different ways -- in the sense that they are not consistent with observed ones -- and yet fall exactly on the observed MDAR. It is thus not enough to check that the MDAR is satisfied.
For example, a simulated galaxy can have a range of radii where accelerations are high and baryons dominate, but with an unacceptable baryon distribution, and so a rotation curve that is unacceptable. These regions would still lie exactly on the Newtonian branch of the MDAR.
\par
This is indeed the case with half (nine) of the galaxies in the simulated sample of Ref. \cite{kw16}.
To quote Ref. \cite{kwc16} from the caption to their Fig. 4, referring to these galaxies:
``As is clear from the above rotation curves, galaxies
which overproduce stars (shown in red) also have large central
concentrations, giving steeply peaked rotation curves inconsistent
with those seen in local L* galaxies.''  Why then, do they think that these galaxies are at all relevant to their MDAR, if they are inconsistent with observed galaxies?
\par
These galaxies are ``wrong'' in that they show very steeply declining rotation curves within the inner $10-20 \kpc$.
For example, from $\sim 700\kms$ at $1\kpc$ to $\sim 350\kms$ at $10\kpc$, or from $\sim 650\kms$ at $1\kpc$ to $\sim 300\kms$ at $10\kpc$.
Observed disc galaxies do not behave like that at all. (Compare, for example, with the observed rotation curves of galaxies in the same mass range shown in Ref. \cite{sn07}.)
\par
These ``unphysical'' regions are easily seen to be of high acceleration, $g\gg\az$, and to be dominated by baryons. Namely, they must all lie tightly on the Newtonian branch of the observed MDAR.
In fact, looking at the rotation curves immediately convinces us that most of the weight on the simulated, Newtonian branch of of Ref. \cite{kw16} comes from these `wrong'' regions;\footnote{See sec. \ref{comments}} this thus happens for a trivial, but wrong reason.
\section{\label{comments}Additional comments}
Here I make some further comments.
\par
Reference \cite{kw16} artificially picks what they call ``data points'' -- better called ``sampling points'' -- across the simulated galaxies, at different radii, plots them in the $g-\gN$ plane, and makes judgements (on scatter, etc.) according to their distribution in that plane. Clearly, the choice of points is arbitrary. But the weight and the scatter around the simulated MDAR can be affected by the choice of these ``data points''. For example, if regions that depart from the observed MDAR are given sparse sampling their weight in the assessment, which relies on the density of data points in the plane, will be low.
\par
In general, the sampling might not reflect faithfully the distribution of the data that goes into the observed relation. For example, it may be that the observed rotation curves do not go very far in radius, for observational reasons, while in the simulations one can sample to very large radii -- since everything can be `observed' there. This might give undue weight to regions far on the asymptotic rotation curve -- which, as we saw, can `trivially' reduce the scatter of the simulated MDAR -- while the real data sample better the inner parts.

\par
So it is not quite clear what meaning to attach to statistical aspects of the results.
\par
The simulated galaxies lie on the observed MDAR only for the present epoch in their evolution (redshift $z=0$).
To quote the caption of Fig. 2 of Ref. \cite{kw16} ``As this figure shows, at higher redshift the
low $g_{bar}$ (my $\gN$) slope is much shallower than at $z=0$''.
In other words it is only today that $g\propto \gN^{1/2}$ for asymptotic rotation curves.  So the present slope of 1/2 is just some coincidence occurring at $z=0$. In, MOND this particular slope is highly significant, as it follows from the space-time scale invariance of the deep-MOND limit.
\par
The simulation  in question attempt to treat very complicated, haphazard, and unknowable events and processes taking place during the formation and evolution histories of these galaxies. The crucial baryonic processes, in particular, are impossible to tackle by actual, true-to-nature, simulation. So they are represented in the simulations by various effective prescriptions, which have many controls and parameters, and which leave much freedom to adjust the outcome of these simulations. Thus for example, results for the same simulated galaxies may vary greatly from one simulation to the next, depending on the way the very-little-understood baryonic physics is treated (see, e.g.,
Ref. \cite{kwc16} for examples).
\par
The exact strategies involved are practically impossible to pinpoint by an outsider, and they probably differ among simulations. But, one will not be amiss to suppose that over the years, the many available handles have been turned so as to get galaxies as close as possible to observed ones.

\end{document}